\documentclass{article}
\usepackage{graphicx}
\usepackage{amsmath}
\usepackage{multirow}
\usepackage{listings}
\usepackage{color}

\definecolor{dkgreen}{rgb}{0,0.6,0}
\definecolor{gray}{rgb}{0.5,0.5,0.5}
\definecolor{mauve}{rgb}{0.58,0,0.82}

\begin{document}
\title{Systematization of Knowledge and Implementation: Short Identity-Based Signatures.}
%
%
\author{Diptendu Mohan Kar and Indrajit Ray \\
Colorado State University, Fort Collins CO 80523, USA \\
\{diptendu.kar, indrajit.ray\}@colostate.edu}

\date{}
\maketitle              
\begin{abstract}
Identity-Based signature schemes are gaining a lot of popularity every day. Over the last decade, there has been a lot of schemes that have been proposed. Several libraries are there that implement identity-based cryptosystems that include identity-based signature schemes like the JPBC library which is written in Java and the charm-crypto library written in python. However, these libraries do not contain all of the popular schemes, rather the JPBC library contains only one identity-based signature scheme and the charm-crypto contains three. Furthermore, the implemented schemes are designed to work on one particular pairing curve. In pairing-based cryptosystems, even for a given signature scheme, the size of the signature and the performance i.e. the time to sign and verify depends on the chosen pairing curve. There are many applications in which the signature size is of more importance than the performance and similarly other applications where the performance is of more importance than signature size. In this work, we describe the popular signature schemes and their implementation using the JPBC library and describe how different pairing curves affect the signature size and performance. We also provide two methods to further shorten the signature size which is not present in the libraries by default.

\end{abstract}

\section{Introduction}
\label{sec:intro}

Identity-based cryptosystems were first proposed by Shamir in 1984\cite{shamir_identity-based_1984}. In his pioneering work, although he came up with a working solution for identity-based signatures, he could not describe a fully working scheme for identity-based encryption. The identity-based encryption is possible only through pairing-based cryptography. Not only identity-based encryption but identity-based key exchange, identity-based identification, and many other schemes are possible because of pairing-based cryptography.
Pairing-based cryptography emerged in 2000 with Joux's tripartite Diffie Hellman\cite{joux} and Sakai's pairing-based cryptosystem\cite{sakai2000}. But it gained a lot of popularity with Dan Boneh's groundbreaking solution to a long time unsolved problem of ``Identity-based Encryption''\cite{boneh2001identity}

Pairing-based cryptography is an extension of elliptic curve cryptography. Pairing-based cryptography uses specially chosen curves that allows us to check more complicated equations over elliptic curve points. Pairing or bilinear pairing (and hence the name) can be thought of as an operation which when applied to elements of two groups, generates an output element which belongs to a third group.  A detailed explanation of how pairing-based cryptography works can be found here \cite{pbcintro}. There are many resources on the internet apart from the referred paper which can be used since this is one of the hot topics in cryptography. We will not be going through all the details of the working principles of pairing-based cryptography. We will be going through the definition of bilinear maps, and then move on to describe the special curves that support pairing. 

The following definition of ``pairing'' is commonly used in most academic papers\cite{pbcintro} - 

Let $G_{1},G_{2}$ be two additive cyclic groups of prime order $q$, and $G_{T}$ another cyclic group of order $q$ written multiplicatively. A pairing is a map: $e: G_{1}\times G_{2}\rightarrow G_{T} $ , which satisfies the following properties:

\textbf{Bilinearity: } $ \forall a,b\in F_{q}^{*}, \forall P\in G_{1}, Q\in G_{2}: e\left(aP,bQ\right)=e\left(P,Q\right)^{ab}$

\textbf{Non-degeneracy: } $e\neq 1$

\textbf{Computability: } There exists an efficient algorithm to compute $e$.

If the same group is used for the first two groups (i.e. $G_{1}=G_{2}$), the pairing is called \textit{symmetric} and is a mapping from two elements of one group to an element from a second group.

Some researchers classify pairing instantiations into three (or more) basic types:
\begin{enumerate}
\item $G_{1}=G_{2}$
\item $G_{1}\neq G_{2}$ but there is an efficiently computable homomorphism $\phi :G_{2}\to G_{1}$.
\item $G_{1}\neq G_{2}$ and there are no efficiently computable homomorphisms between $G_{1}$ and $G_{2}$.
\end{enumerate}

There exist many libraries in different programming languages that implement pairing based cryptosystems and identity-based schemes like identity-based encryption, identity-based signatures, identity-based identification, etc. The very first library which implemented was the PBC library written by Ben Lynn \cite{lynn2006pairing} in C. Using that library similar pairing based libraries have been written in other programming languages like Java - JPBC library \cite{jpbc}and Python - charm crypto library \cite{charm13}. There are many identity-based signature schemes, but in the JPBC library, there is only one scheme that is implemented - Paterson and Schuldt\cite{paterson2006efficient}. Charm crypto library has three IBS schemes implemented. In this work, we are interested in identity-based signatures specifically and find which scheme offers the shortest signature size using the JPBC library. 

There exist many applications where the signature size is of more importance rather than the performance. It is always possible to reduce the signature size just by shortening the length of the prime but that will compromise the security. The curve parameters we will be using are provided in the library already and are considered to be secure. One of the emerging areas that short signatures are applicable is in biology. Kar et. al \cite{nspw} proposed applying digital signatures to the DNA in living organisms. In their work, it is very important to use short signatures as longer signatures might make the organism unstable. In their work, they have used the original Shamir's IBS scheme and 1024 bit keys, but pairing based scheme will be a better choice because we can get better security at much fewer bits. In a pairing based system, even for a fixed signature scheme, the signature size is dependent on the type of curve used for pairing. To the best of our knowledge, there is no other work that talks about the different signature schemes and how they compare using the different curves that are built into the JPBC library. The ultimate goal of this work is to integrate all the schemes with the JPBC library which will provide users with a lot of more choices and flexibility.

The rest of the paper is organized as follows. In section \ref{sec:pairingcurves}, we describe the curve parameters and how each group element will act for each curve. In section \ref{sec:signatureschemes}, we discuss a couple of signature schemes that are very popular yet not implemented in the JPBC library or charm crypto library. In section \ref{sec:shortersig} we show two techniques that can be used to further shorten the signature size which is not built into the libraries. 
\section{Pairing Curves}
\label{sec:pairingcurves}

Pairing curves are the special elliptic curves that support the bilinear pairing operation. In the libraries, they are predefined. There are a couple of ``types" of curves that are present. Each type of curve is different in its base form or equation that defines it. Other than the predefined curves that are already present in the library, the parameters for each curve can be regenerated as to choose better security e.g. length of prime. In the JPBC library, they are present in the ``properties'' files and in the charm crypto library they are present in the ``pairingcurves.py" file. The different types of curves that are available in the JPBC library are as follows:

\textbf{Type A: }Type A pairings are constructed on the curve $y^2 = x^3 + x$ over the field $F_q$ for some prime $q = 3 \;mod \;4$. Both $G_1$ and $G_2$ are the group of points $E(F_q)$, so this pairing is symmetric. It turns out $\#E(F_q) = q + 1$ and $\#E(F_{q^2}) = (q + 1)^2$. Thus the embedding degree $k$ is 2, and hence $G_T$ is a subgroup of $F_{q^2}$. The order $r$ is some prime factor of $q + 1$.

There is only one built-in type A curve in the JPBC library. The parameters in the file \textbf{a.properties} are - 

type a

$q$ 8780710799663312522437781984754049815806883199414208211028653\\3992664756308
8022295707862517942266222142315585876958231745927\\7713367317481324925129998224791

$h$ 1201601226489114607938882136674053420480295440125131182291961\\5131047207289359704531102844802183906537786776

$r$ 730750818665451621361119245571504901405976559617

exp2 159

exp1 107

sign1 1

sign0 1

Here, $q$ is 512 bits or 64 bytes, $r$ is 160 bits or 20 bytes. It is recommended to use at least 512-bit primes for type A curves. Other parameters can be generated using the following line of code - 

ParametersGenerator pbcPg = new PBCTypeACurveGenerator(rBits, qBits);

The size of the group elements are as follows - since $r$ is 20 bytes, any element in the group $Z_r$ is 20 bytes. An element in the group $G_1$ is an element over $F_q$. Since $q$ is 512 bits or 64 bytes, an element of the group $G_1$ is 2*64 = 128 bytes (each element is a point whose x cord is 64 bytes and y cord is 64 bytes). Since this curve uses symmetric pairing, the elements of the group $G_2$ and $G_T$ are also 128 bytes each. In the charm crypto library this same curve is present and is called using the tag ``SS512".

\textbf{Type A1: }Type A1 uses the same equation, but have different fields. It supports fields of composite order. There is only one built-in type A1 curve in the JPBC library. We are not listing the parameters here but it can be found in the \textbf{a1.properties} file. The same curve is present in the charm crypto library and is called using the tag ``SS1024". The prime $n$ has length 1022 bits. The size of the group elements are as follows - any element in the group $Z_r$ is 1024 bits or 128 bytes. An element in the group $G_1$, $G_2$ and $G_T$ are all 2080 bits or 260 bytes. The code to generate type A1 parameters is -
 
ParametersGenerator pbcPg = new PBCTypeA1CurveGenerator();

\textbf{Type D: }These are ordinary curves of with embedding degree 6, whose orders are prime or a prime multiplied by a small constant. A type D curve is defined over some field $F_q$ and has order $h * r$ where $r$ is a prime and $h$ is a small constant. Over the field, $F_{q^6}$ its order is a multiple of $r^2$. Typically the order of the curve $E$ is around 170 bits, as is $F_q$, the base field, thus $q^k$ is around the 1024-bit mark which is commonly considered good enough. These type of curves were discovered by Miyaji, Nakabayashi, and Takano and commonly known as MNT curves\cite{miyaji2001new}.

There are 3 type D curves built-in the library - d159, d201, and d224. The numbers represent the bits in the prime $q$. Let us look at the parameters in the \textbf{d159.properties} file - 

type d

$q$ 625852803282871856053922297323874661378036491717

$n$ 625852803282871856053923088432465995634661283063

$h$ 3

$r$ 208617601094290618684641029477488665211553761021

$a$ 581595782028432961150765424293919699975513269268

$b$ 517921465817243828776542439081147840953753552322

$k$ 6

Here, $q$ is 160 bits or 20 bytes, $r$ is 160 bits or 20 bytes. The parameters $a$ and $b$ are the coefficients in the equation  $E: y^2 = x^3 + ax + b$. The size of the group elements are as follows - since $r$ is 20 bytes, any element in the group $Z_r$ is 20 bytes. An element in the group $G_1$ is an element over $F_q$. Since $q$ is 160 bits or 20 bytes, an element of group $G_1$ is 2*20 = 40 bytes. The elements of the group $G_2$ are over $F_{q^6}$ and hence they are 120 bytes. An element of $G_T$ is also 120 bytes. Apart from the parameters listed above there are some more which we did not show, they are $nk$ -  number of points in $E(F_{q^k})$, $hk$ where $nk = hk * r * r$, coeff0, coeff1, coeff2 and nqr - quadratic nonresidue in $F_q$. The other two files have the same parameters just the prime bits are more. The code to generate type D parameters is -

ParametersGenerator pbcPg = new \\
PBCTypeDParametersGenerator(discriminant);

The charm crypto library contains the same three ``d-type" curves and are called using the tags ``MNT159", ``MNT201" and ``MNT224" respectively.

\textbf{Type E: }The CM (Complex Multiplication) method of constructing elliptic curves starts with the Diophantine equation  $DV^2=4q-t^3$. If  $t$=2 and $q=Dr^2h^2+1$ for some prime $ r$ (which we choose to be a Solinas prime) and some integer  $h$ , we find that this equation is easily solved with $V = 2rh$ .

As $q$ is typically 1024 bits, group elements take a lot of space to represent. Moreover, many optimizations do not apply to this type, resulting in a slower pairing. Hence this type E curve is not generally used in any schemes and so we will not be listing the parameters here. An interested user can look up the file \textbf{e.properties} in the jPBC library. This type is not present in the charm crypto library.

\textbf{Type F: }Using carefully crafted polynomials, k = 12 pairings can be constructed. Only 160 bits are needed to represent elements of one group and 320 bits for the other. Also, embedding degree k = 12 allows higher security short signatures. The equation is of the form $E: y^2 = x^3 + b$. These type of curves were discovered by Barreto and Naehrig and commonly known as BN curves\cite{barreto}.

There is only one type F curve built in the library. The parameters in the file \textbf{f.properties} are - 

type f

$q$ 205523667896953300194896352429254920972540065223

$r$ 205523667896953300194895899082072403858390252929

$b$ 40218105156867728698573668525883168222119515413

beta 115334401956802802075595682801335644058796914268

alpha0 191079354656274778837764015557338301375963168470

alpha1 71445317903696340296199556072836940741717506375

Here, $q$ is 158 bits or 20 bytes, $r$ is 158 bits or 20 bytes. The parameter $b$ is the coefficient in the equation  $E: y^2 = x^3 + b$. The size of the group elements are as follows - since $r$ is 20 bytes, any element in the group $Z_r$ is 20 bytes. An element in the group $G_1$ is an element over $F_q$. Since $q$ is 20 bytes, an element of group $G_1$ is 2*20 = 40 bytes. The elements of the group $G_2$ are 80 bytes and the elements of group $G_T$ are 240 bytes. The code to generate type F parameters is -

ParametersGenerator pbcPg = new PBCTypeFCurveGenerator(rBits);

In the charm crypto library,  a different type f curve is present. In there, $q$ is 254 bits or 32 bytes, $r$ is also 254 bits or 32 bytes. The size of the group elements are as follows - since $r$ is 32 bytes, any element in the group $Z_r$ is 32 bytes. An element in the group $G_1$ is an element over $F_q$. Since $q$ is 32 bytes, an element of group $G_1$ is 2*32 = 64 bytes. The elements of the group $G_2$ are 128 bytes and the elements of group $G_T$ are 384 bytes.

\textbf{Type G: }Type G curves are exactly the same structure as type D curves with the only difference being in the embedding degree. The embedding degree $k$ is of 10. These types of curves were discovered by Freeman\cite{freeman2006constructing}. The parameters are exactly the same as described in type D curve parameters. There is only one type G curve built in the library called \textbf{g149.properties}. The parameters are - 

type g

$q$ 503189899097385532598615948567975432740967203

$n$ 503189899097385532598571084778608176410973351

$h$ 1

$r$ 503189899097385532598571084778608176410973351

$a$ 465197998498440909244782433627180757481058321

$b$ 463074517126110479409374670871346701448503064

$k$ 10

The prime $q$ is 149 bits the order $r$ is also 149 bits.  The size of the group elements are as follows - since $r$ is 149 bits or 19 bytes, any element in the group $Z_r$ is 19 bytes. An element in the group $G_1$ is an element over $F_q$. Since $q$ is 149 bits or 19 bytes, an element of group $G_1$ is 2*19 = 38 bytes. The elements of the group $G_2$ and $G_T$ are 190 bytes each. Apart from the parameters listed above there are some more which we did not show, they are $nk$ -  number of points in $E(F_{q^k})$, $hk$ where $nk = hk * r * r$, coeff0, coeff1, coeff2 and nqr - quadratic nonresidue in $F_q$. The code to generate type G parameters is -

ParametersGenerator pbcPg = new\\
PBCTypeGParametersGenerator(discriminant);

This type is not present in the charm crypto library.

Table 1 summarizes the different types of curves built in the JPBC library and the different sizes of the group elements for each type of curve.

\begin{table}[h]
\centering
\caption{Sizes of the different group elements for different types of curves in JPBC}

\begin{footnotesize}
\begin{tabular}{|c|c|c|c|c|c|} \hline 

\textbf{Curve Name} & \textbf{\begin{tabular}[c]{@{}c@{}}Bits in \\prime (q)\\ /\\Bits in \\order (r)\end{tabular}} & \textbf{\begin{tabular}[c]{@{}c@{}}Size of \\ element in \\ $Z_r$\end{tabular}} & \textbf{\begin{tabular}[c]{@{}c@{}}Size of \\ element in \\ $G_1$\end{tabular}} & \textbf{\begin{tabular}[c]{@{}c@{}}Size of \\ element in \\ $G_2$\end{tabular}} & \textbf{\begin{tabular}[c]{@{}c@{}}Size of \\ element in \\ $G_T$\end{tabular}} \\ \hline
a.properties        & 512 / 160                                                                               & \begin{tabular}[c]{@{}c@{}}160 bits\\ or 20 bytes\end{tabular}                  & \begin{tabular}[c]{@{}c@{}}1024 bits \\ or 128 bytes\end{tabular}               & \begin{tabular}[c]{@{}c@{}}1024 bits \\ or 128 bytes\end{tabular}               & \begin{tabular}[c]{@{}c@{}}1024 bits \\ or 128 bytes\end{tabular}               \\ \hline
a1.properties       & 1033 / 1022                                                                             & \begin{tabular}[c]{@{}c@{}}1024 bits \\ or 128 bytes\end{tabular}               & \begin{tabular}[c]{@{}c@{}}2080 bits\\ or 260 bytes\end{tabular}                & \begin{tabular}[c]{@{}c@{}}2080 bits\\ or 260 bytes\end{tabular}                & \begin{tabular}[c]{@{}c@{}}2080 bits\\ or 260 bytes\end{tabular}                \\ \hline
d159.properties     & 159 / 158                                                                               & \begin{tabular}[c]{@{}c@{}}160 bits\\ or 20 bytes\end{tabular}                  & \begin{tabular}[c]{@{}c@{}}320 bits \\ or 40 bytes\end{tabular}                 & \begin{tabular}[c]{@{}c@{}}960 bits\\ or 120 bytes\end{tabular}                 & \begin{tabular}[c]{@{}c@{}}960 bits\\ or 120 bytes\end{tabular}                 \\ \hline
d201.properties     & 201 / 181                                                                               & \begin{tabular}[c]{@{}c@{}}184 bits\\ or 23 bytes\end{tabular}                  & \begin{tabular}[c]{@{}c@{}}416 bits\\ or 52 bytes\end{tabular}                  & \begin{tabular}[c]{@{}c@{}}1248 bits\\ or 156 bytes\end{tabular}                & \begin{tabular}[c]{@{}c@{}}1248 bits\\ or 156 bytes\end{tabular}                \\ \hline
d224.properties     & 224 / 224                                                                               & \begin{tabular}[c]{@{}c@{}}224 bits \\ or 28 bytes\end{tabular}                 & \begin{tabular}[c]{@{}c@{}}448 bits \\ or 56 bytes\end{tabular}                 & \begin{tabular}[c]{@{}c@{}}1344 bits\\ or 168 bytes\end{tabular}                & \begin{tabular}[c]{@{}c@{}}1344 bits\\ or 168 bytes\end{tabular}                \\ \hline
e.properties        & 1020 / 160                                                                              & \begin{tabular}[c]{@{}c@{}}160 bits\\ or 20 bytes\end{tabular}                  & \begin{tabular}[c]{@{}c@{}}2048 bits\\ or 256 bytes\end{tabular}                & \begin{tabular}[c]{@{}c@{}}2048 bits\\ or 256 bytes\end{tabular}                & \begin{tabular}[c]{@{}c@{}}1024 bits \\ or 128 bytes\end{tabular}               \\ \hline
f.properties        & 158 / 158                                                                               & \begin{tabular}[c]{@{}c@{}}160 bits\\ or 20 bytes\end{tabular}                  & \begin{tabular}[c]{@{}c@{}}320 bits \\ or 40 bytes\end{tabular}                 & \begin{tabular}[c]{@{}c@{}}640 bits\\ or 80 bytes\end{tabular}                  & \begin{tabular}[c]{@{}c@{}}1920 bits\\ or 240 bytes\end{tabular}                \\ \hline
g149.properties     & 149 / 149                                                                               & \begin{tabular}[c]{@{}c@{}}152 bits\\ or 19 bytes\end{tabular}                  & \begin{tabular}[c]{@{}c@{}}304 bits\\ or 38 bytes\end{tabular}                  & \begin{tabular}[c]{@{}c@{}}1520 bits\\ or 190 bytes\end{tabular}                & \begin{tabular}[c]{@{}c@{}}1520 bits\\ or 190 bytes\end{tabular}                \\ \hline
\end{tabular}
\end{footnotesize}
\label{sizecomp}
\end{table}
\section{Identity-based Signature Schemes}
\label{sec:signatureschemes}
As we mentioned earlier that there are several identity-based signature schemes that are constructed on pairings. The only scheme that is implemented in the JPBC library is the scheme by Paterson and Schuldt \cite{paterson2006efficient}. The three schemes implemented in the charm crypto library are - Cha and Cheon \cite{choon2003identity} scheme, Hess \cite{hess2002efficient} scheme, and Waters \cite{waters2005efficient} scheme.  We will describe some notable signature schemes apart from the ones already present in the libraries. Almost every identity-based signature scheme has 4 stages - setup, extract, sign, and verify. In order to have a consistent description of the schemes, we will use a fixed set of notations. 

\textbf{Setup: } This step is executed by the central authority or the PKG. The setup generates the curve parameters. The different curves provided in the library can be used to load the parameters. Let $g_1$ be the generator of $G_1$, $g_2$ be the generator of $G_2$. A random $x \in Z_n^{*}$ is chosen to be the master secret. Two public keys $P_1$ and $P_2$ are calculated as - $P_1 = x \cdot g_1$ and $P_2 = x \cdot g_2$. They are also called master public keys. An embedding function $H$ is chosen such that $H(0,1)^{*} \rightarrow G_1 $. This embedding function converts a string to an element in group $G_1$. These are the public parameters - $<g_1,g_2,P_1,P_2,H>$ and shared with every participant in the system. 

\textbf{Extract: }Takes as input the curve parameters, the master secret key $x$, and a user's identity and returns the users identity-based secret key. This step is performed the central authority for each user $A$ with identity $ID_A$. Every user connects to the central authority over a secure channel and shares its identity. The identity which is mostly a string is converted to an element in $G_1$ using the embedding function $H$. The string is hashed using a hash function like SHA-256, then converted to a number. Then point multiplication is used to get the result - $hash(number) \cdot g_1$. We will call this $C_A$ where $C_A = H(ID_A)$.The syntax using JPBC is - 

String identity = ``myidentitystring";

byte[] idhash = hashSHA256(identity);

Element idelem = G1.newElement().\\
setFromHash(idhash, 0, idhash.length).getImmutable();

Element skid = idelem.powZn(secretKey).getImmutable();

This step can be done by anyone in the system. The users identity-based secret key is calculated as  $V_A = x \cdot C_A$. This can only be done by the PKG. The syntax is -

Element skid = idelem.powZn(secretKey).getImmutable();

These two values - idelem and skid are $C_A$ and $V_A$ respectively. This is returned to the user after the extract phase. Both values are elements of group $G_1$.

\textbf{Sign: }This step is performed by each user who wants to sign. If user A wants to send a message to user B, the first step is to send its identity $ID_A$ to the PKG and get the values $(C_A, V_A)$. Using these values he can proceed to sign the message. The particular steps for signing are different for each scheme as we will show below. 

\textbf{Verify: }This step is performed by the user who received a message. If user A sends a message to user B, B does not need to contact the PKG to get anything (assuming the public parameters are obtained). User B already knows A's identity $ID_A$ as this is public. B also received a message from A with a signature - $M$ and $Sign$. B will invoke verification using $ID_A$, $M$, $Sign$, and public parameters.

We will now describe some well-known signature schemes - 

\subsection{Sakai-Ohgishi-Kasahara scheme}
The Sakai-Ohgishi-Kasahara scheme \cite{sakaicryptosystems} was one of the first identity-based signature schemes that were published in 2000. The steps for generating and verifying the signature are as follows - 

\textbf{Sign: }The user A has already obtained $(C_A,V_A)$ from PKG. To sign a message $m$, a user $A$ with the curve parameters and the secret key $(C_A,V_A)$ does the following:
\begin{enumerate}
\item Choose a random $r \in Z_n^{*}$. Compute $Z_A = r \cdot g_2$. 
\item Compute $R$ by embedding the message $m$ to $G_1$.
\item Compute $S = V_A + rR$. 
\end{enumerate}
A's signature for the message $m$ is - ($Z_A,S$)

\textbf{Verify: }
\begin{enumerate}
\item Compute  $lhs = e_n(S,g_2)$, where $e_n$ is the pairing operation. 
\item Compute $rhs =  e_n(C_A,P_2) * e_n(R,Z_A)$
\item  Check $lhs \stackrel{?}{=} rhs$.
\end{enumerate}

The above equation works because:
\begin{align*}
rhs &= e_n(C_A,P_2) * e_n(R,Z_A) \\
&= e_n(C_A,x \cdot g_2) * e_n(R,r \cdot g_2)\\
&= e_n(x \cdot C_A,g_2) * e_n(r\cdot R,g_2)\\
&= e_n(g_1,g_2)^{x \cdot C_A + r\cdot R}\\
&= e_n(g_1,g_2)^{V_A + r\cdot R}\\
lhs &= e_n(S,g_2)\\
&= e_n((V_A + r.R) \cdot g_1,g_2) \\
&= e_n(g_1,g_2)^{V_A + r\cdot R}
\end{align*}

The signature in this scheme is a tuple $(Z_A,S)$. Note that $Z_A$ is an element of $G_2$ and $S$ is in $G_1$(created from $V_A$ which is in $G_1$ and $R$ which is also in $G_1$). Using the table \ref{sizecomp}, we can see the signature size for the different curves are - 
\begin{table}[]
\centering
\caption{Signature size using different curves for the
  Sakai-Ohgishi-Kasahara scheme}
\label{tab:my-table}
\begin{footnotesize}
\begin{tabular}{|c|c|c|c|c|c|c|c|}
\hline
\textbf{Curve type}                                                       & \textbf{\;a\;} & \textbf{\;a1\;} & \textbf{ d159 } & \textbf{ d201 } & \textbf{ d224 } & \textbf{\;f\;} & \textbf{\;g\;} \\ \hline
\textbf{\begin{tabular}[c]{@{}c@{}}Signature size\\ (bytes)\end{tabular}} & \;256\;        & \;520\;         & 160           & 208           & 224           & \;120\;        & \;228\;        \\ \hline
\end{tabular}
\end{footnotesize}
\end{table}

\subsection{Paterson scheme}
The Paterson scheme \cite{paterson2002id}, 2002 is another notable signature scheme . The steps for generating and verifying the signature are as follows - 

\textbf{Sign: }The user A has already obtained $(C_A,V_A)$ from PKG. To sign a message $m$, a user $A$ with the curve parameters and the secret key $(C_A,V_A)$ does the following:
\begin{enumerate}
\item Choose a random $r \in Z_n^{*}$. Compute $Z_A = r \cdot g_2$. 
\item Compute the hash $h_0$ of the message $m$ using any standard hash function like SHA-256. $h_0 = H(m)$.
\item Compute the hash of $Z_A$ using any standard hash function like SHA-256. $h_1 = H(Z_A)$
\item Compute $S = r^{-1} \cdot h_0 \cdot g_1 + r^{-1} \cdot h_1 \cdot V_A$
\end{enumerate}
A's signature for the message $m$ is - ($Z_A,S$)

\textbf{Verify: }
\begin{enumerate}
\item Compute  $lhs = e_n(S,Z_A)$, where $e_n$ is the pairing operation. 
\item Compute $rhs =  e_n(g_1,g_2)^{h_0} * e_n(C_A,P_2)^{h_1}$
\item  Check $lhs \stackrel{?}{=} rhs$.
\end{enumerate}
The above equation works because:
\begin{align*}
rhs &= e_n(g_1,g_2)^{h_0} * e_n(C_A,P_2)^{h_1} \\
&= e_n(g_1,g_2)^{h_0} * e_n(C_A,x \cdot g_2)^{h_1}\\
&= e_n(g_1,g_2)^{h_0} * e_n(x \cdot C_A, g_2)^{h_1}\\
&= e_n(g_1,g_2)^{h_0} * e_n(V_A, g_2)^{h_1}\\
&= e_n(g_1,g_2)^{h_0 + V_A * h_1}\\
lhs &= e_n(S,Z_A)\\
&= e_n(r^{-1} \cdot h_0 \cdot g_1 + r^{-1} \cdot h_1 \cdot V_A, r \cdot g_2) \\
&= e_n(r^{-1}\cdot (h_0 + h_1 * V_A) \cdot g_1, r \cdot g_2)\\
&= e_n(g_1,g_2)^{r^{-1} \cdot (h_0 + V_A * h_1) \cdot r}\\
&= e_n(g_1,g_2)^{h_0 + V_A * h_1}
\end{align*}

The signature in this scheme is also a tuple $(Z_A,S)$. Note that $Z_A$ is an element of $G_2$ and $S$ is in $G_1$(created from $V_A$ which is in $G_1$ and $g_1$ which is also in $G_1$). This scheme generates the same signature size as the Sakai-Ohgishi-Kasahara scheme but the construction is different. Using the table \ref{sizecomp}, we can see the signature size for the different curves are - 
\begin{table}[]
\centering
\caption{Signature size using different curves for the Paterson scheme}
\label{tab:my-table}
\begin{footnotesize}
\begin{tabular}{|c|c|c|c|c|c|c|c|}
\hline
\textbf{Curve type}                                                       & \textbf{\;a\;} & \textbf{\;a1\;} & \textbf{ d159 } & \textbf{ d201 } & \textbf{ d224 } & \textbf{\;f\;} & \textbf{\;g\;} \\ \hline
\textbf{\begin{tabular}[c]{@{}c@{}}Signature size\\ (bytes)\end{tabular}} & \;256\;        & \;520\;         & 160           & 208           & 224           & \;120\;        & \;228\;        \\ \hline
\end{tabular}
\end{footnotesize}
\end{table}

\subsection{Sakai-Kasahara scheme}
The Sakai-Kasahara scheme \cite{sakai2003id} was published in 2003. In their work, the authors described two types of signature schemes - one they called El-Gamal analog and the other Schnorr analog. The construction of the El-Gamal analog is as follows- 

\textbf{Sign: }The user A has already obtained $(C_A,V_A)$ from PKG. To sign a message $m$, a user $A$ with the curve parameters and the secret key $(C_A,V_A)$ does the following:
\begin{enumerate}
\item Choose a random $r \in Z_n^{*}$. Compute $Z_A = r \cdot g_2$. 
\item Choose $x_{za}$ as the x-coordinate of $Z_A$.
\item Computer the hash of the message $m$ using any standard hash function like SHA-256. $h = H(m)$.
\item Compute the hash of $Z_A$ using any standard hash function like SHA-256. $h_1 = H(Z_A)$
\item Compute $S = r^{-1} \cdot h \cdot C_A + r^{-1} \cdot x_{za} \cdot V_A$
\end{enumerate}
A's signature for the message $m$ is - ($Z_A,S$)

\textbf{Verify: }
\begin{enumerate}
\item Compute  $lhs = e_n(S,Z_A)$, where $e_n$ is the pairing operation. 
\item Compute $rhs =  e_n(C_A,h \cdot g_2 + x_{za} \cdot P_2)$
\item  Check $lhs \stackrel{?}{=} rhs$.
\end{enumerate}
The above equation works because:
\begin{align*}
lhs &= e_n(S,Z_A) \\
&= e_n(r^{-1} \cdot h \cdot C_A + r^{-1} \cdot x_{za} \cdot V_A,r \cdot g_2)\\
&= e_n(r^{-1} \cdot h \cdot C_A + r^{-1} \cdot x_{za} \cdot x \cdot C_A,r \cdot g_2)\\
&= e_n(r^{-1} \cdot (h + x_{za} \cdot x) \cdot C_A ,r \cdot g_2)\\
&= e_n(C_A,g_2)^{r^{-1} \cdot (h + x_{za} \cdot x) \cdot r }\\
& = e_n(C_A,g_2)^{(h + x_{za}\cdot x) }\\
rhs &= e_n(C_A,h \cdot g_2 + x_{za} \cdot P_2)\\
&= e_n(C_A,h \cdot g_2 + x_{za} \cdot x \cdot g_2)\\
&= e_n(C_A,(h + x_{za} \cdot x) g_2)\\
& = e_n(C_A,g_2)^{(h + x_{za}\cdot x) }\\
\end{align*}
The signature in this scheme is also a tuple $(Z_A,S)$. Note that $Z_A$ is an element of $G_2$ and $S$ is in $G_1$(created from $V_A$ which is in $G_1$ and $C_A$ which is also in $G_1$). This scheme generates the same signature size as the two previous schemes but the construction is different. Using the table \ref{sizecomp}, we can see the signature size for the different curves are - 
\begin{table}[]
\centering
\caption{Signature size using different curves for the Sakai-Kasahara
  scheme - El-Gamal Analogue}
\label{tab:my-table}
\begin{footnotesize}
\begin{tabular}{|c|c|c|c|c|c|c|c|}
\hline
\textbf{Curve type}                                                       & \textbf{\;a\;} & \textbf{\;a1\;} & \textbf{ d159 } & \textbf{ d201 } & \textbf{ d224 } & \textbf{\;f\;} & \textbf{\;g\;} \\ \hline
\textbf{\begin{tabular}[c]{@{}c@{}}Signature size\\ (bytes)\end{tabular}} & \;256\;        & \;520\;         & 160           & 208           & 224           & \;120\;        & \;228\;        \\ \hline
\end{tabular}
\end{footnotesize}
\end{table}

The Sakai-Kasahara Schnorr analogue construction is as follows:

\textbf{Sign: }To sign a message $m$, a user $A$ with the curve parameters and the secret key $(C_A,V_A)$ does the following:
\begin{enumerate}
\item Choose a random $r \in Z_n^{*}$. Compute $Z_A = r \cdot g_2$. 
\item Compute $e = e_n(C_A,Z_A)$, where $e_n$ is the pairing operation.
\item Compute $h = H_1(m \parallel e)$, where $H_1$ is a secure cryptographic hash function such as SHA-256 and $\parallel$ is the concatenation operation.
\item Compute $S = hV_A + rC_A$ .
\end{enumerate}
A's signature for the message $m$ is - ($h,S$)

\textbf{Verify: } The verification procedure is as follows:
\begin{enumerate}
\item Compute $w = e_n(S,g_2) * e_n(C_A, - h P_2 )$
\item Check $H_1(m \parallel w) \stackrel{?}{=} h$
\end{enumerate}

The above equation works because:
\begin{align*}
e &= e_n(C_A,Z_A) = e_n(C_A,r\cdot g_2) = e_n(C_A,g_2)^r\\
w &= e_n(S,g_2) * e_n(C_A, - h P_2 )\\
 &= e_n( hV_A + rC_A, g_2) *  e_n(C_A, - h x \cdot g_2)\\
 &=  e_n( h x \cdot C_A + rC_A, g_2) *  e_n(C_A, g_2)^{- h x} \\
 &=  e_n( (h x + r) \cdot C_A, g_2)  *  e_n(C_A, g_2)^{- h x} \\
 &=  e_n(C_A, g_2)^{h x + r} *  e_n(C_A, g_2)^{- h x} \\
 &=   e_n(C_A, g_2)^{r}
\end{align*}

The signature in this scheme is also a tuple $(h,S)$. Here $h$ is a hash function. If we used SHA-160 then the size of $h$ would be 20 bytes but it is not recommended nowadays. For SHA-256 the size of $h$ is 32 bytes. The parameter $S$ is an element of group $G1$ Using the table \ref{sizecomp}, we can see the signature size for the different curves are - 
\begin{table}[]
\centering
\caption{Signature size using different curves for the Sakai-Kasahara
  scheme - Schnorr Analogue}
\label{tab:my-table}
\begin{footnotesize}
\begin{tabular}{|c|c|c|c|c|c|c|c|}
\hline
\textbf{Curve type}                                                       & \textbf{\;a\;} & \textbf{\;a1\;} & \textbf{ d159 } & \textbf{ d201 } & \textbf{ d224 } & \textbf{\;f\;} & \textbf{\;g\;} \\ \hline
\textbf{\begin{tabular}[c]{@{}c@{}}Signature size\\ (bytes)\end{tabular}} & \;160\;        & \;292\;         & 72           & 84           & 88           & \;72\;        & \;70\;        \\ \hline
\end{tabular}
\end{footnotesize}
\end{table}

All of the previous 3 schemes had an element of $G_2$ in their signature which caused the size to grow because an element of $G_2$ is larger than an element of $G_1$ for asymmetric pairings. For symmetric pairings, the element size is the same in both. 

\subsection{Xun-Yi scheme}
The Xun-Yi scheme \cite{yi2003identity}, published in 2003 is another notable signature scheme. The steps for generating and verifying the signature are as follows - 

\textbf{Sign: }The user A has already obtained $(C_A,V_A)$ from PKG. To sign a message $m$, a user $A$ with the curve parameters and the secret key $(C_A,V_A)$ does the following:
\begin{enumerate}
\item Choose a random $r \in Z_n^{*}$. Compute $Z_A = r \cdot g_1$. 
\item Compute $h = H_1(m \parallel Z_A)$, where $H_1$ is a secure cryptographic hash function such as SHA-256 and $\parallel$ is the concatenation operation.
\item Compute $S = r \cdot P_1 + h \cdot V_A$
\end{enumerate}
A's signature for the message $m$ is - ($Z_A,S$)

\textbf{Verify: }
\begin{enumerate}
\item Compute  $lhs = e_n(S,g_2)$, where $e_n$ is the pairing operation. 
\item Compute $rhs =  e_n(Z_A + h \cdot C_A, P_2)$
\item  Check $lhs \stackrel{?}{=} rhs$.
\end{enumerate}
The above equation works because:
\begin{align*}
rhs &= e_n(Z_A + h \cdot C_A, P_2)\\
&= e_n(r \cdot g_1 + h \cdot C_A, x \cdot g_2)\\
&= e_n(g_1,g_2)^{(r + h \cdot C_A) \cdot x}\\
lhs &= e_n(S,g_2)\\
&= e_n( r \cdot P_1 + h \cdot V_A,g_2) \\
&= e_n( r \cdot x \cdot g_1 + h \cdot x \cdot C_A,g_2) \\
&= e_n(g_1,g_2)^{(r + h \cdot C_A) \cdot x}\\
\end{align*}

The signature in this scheme is also a tuple $(Z_A,S)$. But both $Z_A$ and $S$ are elements of $G_1$. Using the table \ref{sizecomp}, we can see the signature size for the different curves are - 
\begin{table}[]
\centering
\caption{Signature size using different curves for the Xun Yi scheme.}
\label{tab:my-table}
\begin{footnotesize}
\begin{tabular}{|c|c|c|c|c|c|c|c|}
\hline
\textbf{Curve type}                                                       & \textbf{\;a\;} & \textbf{\;a1\;} & \textbf{ d159 } & \textbf{ d201 } & \textbf{ d224 } & \textbf{\;f\;} & \textbf{\;g\;} \\ \hline
\textbf{\begin{tabular}[c]{@{}c@{}}Signature size\\ (bytes)\end{tabular}} & \;256\;        & \;520\;         & 80           & 104           & 112           & \;80\;        & \;76\;        \\ \hline
\end{tabular}
\end{footnotesize}
\end{table}

Comparing the schemes above, we can observe that the Sakai-Kasahara Schnorr analog and the Xun Yi scheme produces much smaller signature sizes than the other. This is mainly due to its construction. It can be observed that the smaller signatures do not contain any element of $G_2$ in their signature. The Sakai-Kasahara contains a hash and an element of $G_1$ and Xun Yi contains two elements of $G_1$.
\section{Shorter signature size}
\label{sec:shortersig}
In almost all of the signature schemes, the signature of a message contains an element in group $G_1$. As we observed for the schemes described above, the signature is a tuple and one of the variables is from the group $G_1$. The Paterson and Schuldt scheme \cite{paterson2006efficient} which is present in the JPBC library, the signature size is a triple where all the three variables are elements in the group $G_1$. Also, in the Cha-Cheon scheme present in the charm crypto library, the signature is a tuple, where both variables are elements in group $G_1$. We did not describe these two schemes as they are already implemented. Now, the elements in the group $G_1$ are points on the elliptic curve. They have two subcomponents the X-coordinate and the Y-coordinate which are packed together. In the JPBC library, the elements are assigned and computed using the class ``Element". The fields are generated from the pairing curves i.e. the properties files. When we print the ``Element" from the group $G_1$, we can observe that it contains the two coordinates. The following syntax is used to generate a random element in $G_1$ -

Pairing pairing = PairingFactory.getPairing(``f.properties");

PairingFactory.getInstance().setUsePBCWhenPossible(true);

Field G1 = pairing.getG1();

Element g1 = G1.newRandomElement().getImmutable();
        
System.out.println("g1 - "+g1);
        
System.out.println("G1 bytes = "+g1.toBytes().length);

The code produces the following output - 

g1 - 1850050205405678718762488884335150904922997774,\\
55999770258652075328012601245471415805643870392,0\\
G1 bytes = 40

In order to extract the X and Y coordinate separately, the Element class needs to be converted to a byte array. In that array, half of the array contains the X coordinate the other half contains the Y coordinate. We can then separate the two arrays and convert them to the BigInteger class.

Now every elliptic curve and consequently every pairing curve is defined by an equation of the form $y^2 = x^3 + ax + b$. Hence, the Y coordinate can be calculated from the X coordinate. We do not need to store the Y coordinate in the signature. The Y coordinate can be calculated from the X coordinate during verification. But it has to be noted that when we plug in the value of X, there will be two solutions for the Y coordinate and the square root is modulo prime. For this reason, the way to distinguish between which value of Y to keep, there needs to be some more data about the Y coordinate. The convention to do this is to add one extra byte that will denote if the Y coordinate is odd or even. When we discard the Y coordinate we can do a modulo 2 and if Y is odd, we append the byte 02 before X value. If Y is odd we append 03. Using this point compression technique, it can be observed that a signature which contains an element of $G_1$ and has a size of 2$n$ bytes(assuming each X and Y are $n$ bytes), can be compressed to a size of ($n$ + 1) bytes. The code for point compression is provided in the Appendix.

This technique can only be applied to signature schemes where the signature contains an element in $G_1$. We are not sure if the elements of $G_2$ and $G_T$ can be compressed. So for all of the schemes that are implemented already and the schemes that we described above can utilize this compression. Also, the Sakai-Kasahara Schnorr analog scheme contains a hash value in one of the tuple. Recall that the Sakai-Kasahara signature was ($h, S$), where $h$ is a hash function like SHA-256. Along with the point compression, the hash value can also be shortened using the techniques that are used to generate Ethereum\cite{wood2014ethereum} or Bitcoin \cite{nakamoto2008bitcoin} addresses. In Ethereum, the public key is hashed using keccak 256 hash algorithm. Then instead of taking the entire 32 bytes (64 bytes in hex representation), the Ethereum blockchain takes only the last 20 bytes (40 bytes in hex) and generates the wallet address. Each wallet address is 40 hex bytes but the hash generates 64 hex bytes. There is also another way of doing the same hash compression. When computing the verification step - $ e_n(C_A, - h P_2 )$. Here $- h$ is the negative hash value integer and we perform a scalar multiplication with the point $P_2$. There is only one integer group involved and that is $Z_r$ where $r$ is the order of the curve. So when performing that scalar multiplication $-h \cdot P_2$, JPBC internally converts the hash value to an element of $Z_r$ by modulo $r$. Hence, instead of writing the signature as ($h,S$) we can rewrite that as $(R,S)$ where $R = h \; mod \; r$. This implies our signature is now of the form ($R, S$), where $R$ is an element of the group $Z_r$. But this method will not be efficient for curves which have large orders like the elements in the type a1 curve are 128 bytes. 

In table 7, we list the signature sizes for the signature schemes with and without the point compression. For the Sakai-Kasahara Schnorr scheme, the size is inclusive of the hash compression along with the point compression.

\begin{table}[h]
\centering
\caption{Signature size of different schemes with and without compression}
\label{tab:my-table}
\resizebox{\textwidth}{!}{%
\begin{tabular}{|c|c|l|l|l|l|l|l|l|}
\hline
\textbf{Signature Scheme} & \textbf{Curve type} & \multicolumn{1}{c|}{\textbf{a}} & \multicolumn{1}{c|}{\textbf{a1}} & \multicolumn{1}{c|}{\textbf{d159}} & \multicolumn{1}{c|}{\textbf{d201}} & \multicolumn{1}{c|}{\textbf{d224}} & \multicolumn{1}{c|}{\textbf{f}} & \multicolumn{1}{c|}{\textbf{g}} \\ \hline
\multirow{2}{*}{\textbf{\begin{tabular}[c]{@{}c@{}}Sakai\\ -Ohgishi\\ -Kasahara\\ Signature is \\ (G1,G2)\end{tabular}}} & \textbf{\begin{tabular}[c]{@{}c@{}}Signature size\\ without compression\\ (bytes)\end{tabular}} & \multicolumn{1}{c|}{\begin{tabular}[c]{@{}c@{}}128 + 128\\ = 256\end{tabular}} & \multicolumn{1}{c|}{\begin{tabular}[c]{@{}c@{}}260 + 260\\ = 520\end{tabular}} & \multicolumn{1}{c|}{\begin{tabular}[c]{@{}c@{}}40 + 120\\ = 160\end{tabular}} & \multicolumn{1}{c|}{\begin{tabular}[c]{@{}c@{}}52 + 156\\ = 208\end{tabular}} & \multicolumn{1}{c|}{\begin{tabular}[c]{@{}c@{}}56 + 168\\ = 224\end{tabular}} & \multicolumn{1}{c|}{\begin{tabular}[c]{@{}c@{}}40 + 80\\ = 120\end{tabular}} & \multicolumn{1}{c|}{\begin{tabular}[c]{@{}c@{}}38 + 190\\ = 228\end{tabular}} \\ \cline{2-9} 
 & \textbf{\begin{tabular}[c]{@{}c@{}}Signature size\\ with compression\\ (bytes)\end{tabular}} & \multicolumn{1}{c|}{\begin{tabular}[c]{@{}c@{}}65 + 128\\ = 193\end{tabular}} & \multicolumn{1}{c|}{\begin{tabular}[c]{@{}c@{}}131 + 260\\ = 391\end{tabular}} & \multicolumn{1}{c|}{\begin{tabular}[c]{@{}c@{}}21 + 120\\ = 141\end{tabular}} & \multicolumn{1}{c|}{\begin{tabular}[c]{@{}c@{}}27 + 156\\ = 183\end{tabular}} & \multicolumn{1}{c|}{\begin{tabular}[c]{@{}c@{}}29 + 168\\ = 197\end{tabular}} & \multicolumn{1}{c|}{\begin{tabular}[c]{@{}c@{}}21 + 80\\ = 101\end{tabular}} & \multicolumn{1}{c|}{\begin{tabular}[c]{@{}c@{}}20 + 190\\ = 210\end{tabular}} \\ \hline
\multirow{2}{*}{\textbf{\begin{tabular}[c]{@{}c@{}}Paterson\\ Signature is \\ (G1,G2)\end{tabular}}} & \textbf{\begin{tabular}[c]{@{}c@{}}Signature size\\ without compression\\ (bytes)\end{tabular}} & \multicolumn{1}{c|}{\begin{tabular}[c]{@{}c@{}}128 + 128\\ = 256\end{tabular}} & \multicolumn{1}{c|}{\begin{tabular}[c]{@{}c@{}}260 + 260\\ = 520\end{tabular}} & \multicolumn{1}{c|}{\begin{tabular}[c]{@{}c@{}}40 + 120\\ = 160\end{tabular}} & \multicolumn{1}{c|}{\begin{tabular}[c]{@{}c@{}}52 + 156\\ = 208\end{tabular}} & \multicolumn{1}{c|}{\begin{tabular}[c]{@{}c@{}}56 + 168\\ = 224\end{tabular}} & \multicolumn{1}{c|}{\begin{tabular}[c]{@{}c@{}}40 + 80\\ = 120\end{tabular}} & \multicolumn{1}{c|}{\begin{tabular}[c]{@{}c@{}}38 + 190\\ = 228\end{tabular}} \\ \cline{2-9} 
 & \textbf{\begin{tabular}[c]{@{}c@{}}Signature size\\ with compression\\ (bytes)\end{tabular}} & \multicolumn{1}{c|}{\begin{tabular}[c]{@{}c@{}}65 + 128\\ = 193\end{tabular}} & \multicolumn{1}{c|}{\begin{tabular}[c]{@{}c@{}}131 + 260\\ = 391\end{tabular}} & \multicolumn{1}{c|}{\begin{tabular}[c]{@{}c@{}}21 + 120\\ = 141\end{tabular}} & \multicolumn{1}{c|}{\begin{tabular}[c]{@{}c@{}}27 + 156\\ = 183\end{tabular}} & \multicolumn{1}{c|}{\begin{tabular}[c]{@{}c@{}}29 + 168\\ = 197\end{tabular}} & \multicolumn{1}{c|}{\begin{tabular}[c]{@{}c@{}}21 + 80\\ = 101\end{tabular}} & \multicolumn{1}{c|}{\begin{tabular}[c]{@{}c@{}}20 + 190\\ = 210\end{tabular}} \\ \hline
\multirow{2}{*}{\textbf{\begin{tabular}[c]{@{}c@{}}Sakai \\ - Kasahara\\ El-Gamal\\ Signature is\\ (G1,G2)\end{tabular}}} & \textbf{\begin{tabular}[c]{@{}c@{}}Signature size\\ without compression\\ (bytes)\end{tabular}} & \begin{tabular}[c]{@{}l@{}}128 + 128\\ = 256\end{tabular} & \begin{tabular}[c]{@{}l@{}}260 + 260\\ = 520\end{tabular} & \begin{tabular}[c]{@{}l@{}}40 + 120\\ = 160\end{tabular} & \begin{tabular}[c]{@{}l@{}}52 + 156\\ = 208\end{tabular} & \begin{tabular}[c]{@{}l@{}}56 + 168\\ = 224\end{tabular} & \begin{tabular}[c]{@{}l@{}}40 + 80\\ = 120\end{tabular} & \begin{tabular}[c]{@{}l@{}}38 + 190\\ = 228\end{tabular} \\ \cline{2-9} 
 & \textbf{\begin{tabular}[c]{@{}c@{}}Signature size\\ with compression\\ (bytes)\end{tabular}} & \begin{tabular}[c]{@{}l@{}}65 + 128\\ = 193\end{tabular} & \begin{tabular}[c]{@{}l@{}}131 + 260\\ = 391\end{tabular} & \begin{tabular}[c]{@{}l@{}}21 + 120\\ = 141\end{tabular} & \begin{tabular}[c]{@{}l@{}}27 + 156\\ = 183\end{tabular} & \begin{tabular}[c]{@{}l@{}}29 + 168\\ = 197\end{tabular} & \begin{tabular}[c]{@{}l@{}}21 + 80\\ = 101\end{tabular} & \begin{tabular}[c]{@{}l@{}}20 + 190\\ = 210\end{tabular} \\ \hline
\multirow{2}{*}{\textbf{\begin{tabular}[c]{@{}c@{}}Sakai \\ - Kasahara\\ Schnorr\\ Signature is\\ (hash,G1)\end{tabular}}} & \textbf{\begin{tabular}[c]{@{}c@{}}Signature size\\ without compression\\ (bytes)\end{tabular}} & \begin{tabular}[c]{@{}l@{}}32 + 128\\ = 160\end{tabular} & \begin{tabular}[c]{@{}l@{}}32 + 260\\ = 292\end{tabular} & \begin{tabular}[c]{@{}l@{}}32 + 40\\ = 72\end{tabular} & \begin{tabular}[c]{@{}l@{}}32 + 52\\ = 84\end{tabular} & \begin{tabular}[c]{@{}l@{}}32 + 56\\ = 88\end{tabular} & \begin{tabular}[c]{@{}l@{}}32 + 40\\ = 72\end{tabular} & \begin{tabular}[c]{@{}l@{}}32 + 38\\ = 70\end{tabular} \\ \cline{2-9} 
 & \textbf{\begin{tabular}[c]{@{}c@{}}Signature size\\ with compression\\ (bytes)\end{tabular}} & \begin{tabular}[c]{@{}l@{}}20 + 65\\ = 85\end{tabular} & \begin{tabular}[c]{@{}l@{}}20 + 131\\ = 151\end{tabular} & \begin{tabular}[c]{@{}l@{}}20 + 21\\ = 41\end{tabular} & \begin{tabular}[c]{@{}l@{}}20 + 27\\ = 47\end{tabular} & \begin{tabular}[c]{@{}l@{}}20 + 29\\ = 49\end{tabular} & \begin{tabular}[c]{@{}l@{}}20 + 21\\ = 41\end{tabular} & \begin{tabular}[c]{@{}l@{}}20 + 20\\ = 40\end{tabular} \\ \hline
\multirow{2}{*}{\textbf{\begin{tabular}[c]{@{}c@{}}Xun Yi\\ Signature is\\ (G1,G1)\end{tabular}}} & \textbf{\begin{tabular}[c]{@{}c@{}}Signature size\\ without compression\\ (bytes)\end{tabular}} & \begin{tabular}[c]{@{}l@{}}128 + 128\\ = 256\end{tabular} & \begin{tabular}[c]{@{}l@{}}260 + 260\\ = 520\end{tabular} & \begin{tabular}[c]{@{}l@{}}40 + 40\\ = 80\end{tabular} & \begin{tabular}[c]{@{}l@{}}52 + 52\\ = 104\end{tabular} & \begin{tabular}[c]{@{}l@{}}56 + 56\\ = 112\end{tabular} & \begin{tabular}[c]{@{}l@{}}40 + 40\\ = 80\end{tabular} & \begin{tabular}[c]{@{}l@{}}38 + 38\\ = 76\end{tabular} \\ \cline{2-9} 
 & \textbf{\begin{tabular}[c]{@{}c@{}}Signature size\\ with compression\\ (bytes)\end{tabular}} & \begin{tabular}[c]{@{}l@{}}65 + 65\\ = 130\end{tabular} & \begin{tabular}[c]{@{}l@{}}131 + 131\\ = 262\end{tabular} & \begin{tabular}[c]{@{}l@{}}21 + 21\\ = 42\end{tabular} & \begin{tabular}[c]{@{}l@{}}27 + 27\\ = 54\end{tabular} & \begin{tabular}[c]{@{}l@{}}29 + 29\\ = 58\end{tabular} & \begin{tabular}[c]{@{}l@{}}21 + 21\\ = 42\end{tabular} & \begin{tabular}[c]{@{}l@{}}20 + 20\\ = 40\end{tabular} \\ \hline
\multirow{2}{*}{\textbf{\begin{tabular}[c]{@{}c@{}}Cha - Cheon\\ (available in\\ charm crypto)\\ Signature is\\ (G1,G1)\end{tabular}}} & \textbf{\begin{tabular}[c]{@{}c@{}}Signature size\\ without compression\\ (bytes)\end{tabular}} & \begin{tabular}[c]{@{}l@{}}128 + 128\\ = 256\end{tabular} & \begin{tabular}[c]{@{}l@{}}260 + 260\\ = 520\end{tabular} & \begin{tabular}[c]{@{}l@{}}40 + 40\\ = 80\end{tabular} & \begin{tabular}[c]{@{}l@{}}52 + 52\\ = 104\end{tabular} & \begin{tabular}[c]{@{}l@{}}56 + 56\\ = 112\end{tabular} & \begin{tabular}[c]{@{}l@{}}40 + 40\\ = 80\end{tabular} & \begin{tabular}[c]{@{}l@{}}38 + 38\\ = 76\end{tabular} \\ \cline{2-9} 
 & \textbf{\begin{tabular}[c]{@{}c@{}}Signature size\\ with compression\\ (bytes)\end{tabular}} & \begin{tabular}[c]{@{}l@{}}65 + 65\\ = 130\end{tabular} & \begin{tabular}[c]{@{}l@{}}131 + 131\\ = 262\end{tabular} & \begin{tabular}[c]{@{}l@{}}21 + 21\\ = 42\end{tabular} & \begin{tabular}[c]{@{}l@{}}27 + 27\\ = 54\end{tabular} & \begin{tabular}[c]{@{}l@{}}29 + 29\\ = 58\end{tabular} & \begin{tabular}[c]{@{}l@{}}21 + 21\\ = 42\end{tabular} & \begin{tabular}[c]{@{}l@{}}20 + 20\\ = 40\end{tabular} \\ \hline
\multirow{2}{*}{\textbf{\begin{tabular}[c]{@{}c@{}}Paterson \\ - Schuldt\\ (available in JPBC)\\ Signature is \\ (G1,G1,G1)\end{tabular}}} & \textbf{\begin{tabular}[c]{@{}c@{}}Signature size\\ without compression\\ (bytes)\end{tabular}} & \begin{tabular}[c]{@{}l@{}}3 * 128\\ = 384\end{tabular} & \begin{tabular}[c]{@{}l@{}}3 * 260\\ = 780\end{tabular} & \begin{tabular}[c]{@{}l@{}}3 * 40\\ = 120\end{tabular} & \begin{tabular}[c]{@{}l@{}}3 * 52\\ = 156\end{tabular} & \begin{tabular}[c]{@{}l@{}}3 * 56\\ = 168\end{tabular} & \begin{tabular}[c]{@{}l@{}}3 * 40\\ = 120\end{tabular} & \begin{tabular}[c]{@{}l@{}}3 * 38\\ = 114\end{tabular} \\ \cline{2-9} 
 & \textbf{\begin{tabular}[c]{@{}c@{}}Signature size\\ with compression\\ (bytes)\end{tabular}} & \begin{tabular}[c]{@{}l@{}}3 * 65\\ = 195\end{tabular} & \begin{tabular}[c]{@{}l@{}}3 * 131\\ = 393\end{tabular} & \begin{tabular}[c]{@{}l@{}}3 * 21\\ = 63\end{tabular} & \begin{tabular}[c]{@{}l@{}}3 * 27\\ = 81\end{tabular} & \begin{tabular}[c]{@{}l@{}}3 * 29\\ = 87\end{tabular} & \begin{tabular}[c]{@{}l@{}}3 * 21\\ = 63\end{tabular} & \begin{tabular}[c]{@{}l@{}}3 * 20\\ = 60\end{tabular} \\ \hline
\end{tabular}%
}
\end{table}

In order to recover the Y coordinate during verification we need to do point decompression i.e. retrieve Y using X and the extra byte. The X value is plugged in the equation and the square root modulo prime is calculated. The two values of Y are obtained. Then using the extra byte we know if we should keep the odd Y or the even Y. Note that every curve has a different equation so the decompression method needs to consider the curve and the coefficients a and b. Also, there exists a very easy way to compute the square root modulo prime when the prime modulo 4 equals 3. If prime $\equiv$ 3 mod 4 -

$Y coordinate = (Y^2)^{\frac{prime + 1}{4}} \; mod \; prime $.

However if prime $\equiv$ 1 mod 4, then there is no one line way to calculate the square root. The code for point decompression is provided in the Appendix.

\section{Conclusion}
\label{sec:conclusion}
In this work, we described some of the popular identity based signature schemes. A few of them have been implemented. We did not describe the already implemented schemes but in comparison we listed how their signature sizes compare with other schemes which have not been implemented. From the schemes discussed in this work, it can be concluded that the signature scheme that produces the shortest signature size is the Sakai - Kasahara Schnorr type signature. But this scheme has not been implemented yet in either JPBC or charm crypto library. We also discussed techniques to further shorten the signature size which is not present in either of the libraries yet. 

In the work by Kar \cite{nspw}, they have used Shamir's IBS and using 1024 bit key they generated 1024 bit signature size which translates to 128 bytes. When encoding the signature as nucleotide bases i.e. A, C, G, and T, the length becomes 512 base pairs (each base pair can be written in 2 bits). Using the Sakai- Kasahara scheme along with the compression technique, the signature can be reduced to 41 bytes or 164 base pairs using the type f or d159 curve and to 40 bytes or 160 base pairs using the g type curve.
 
Although we can do point compression on elements of $G1$, it will be interesting to see if the elements of $G2$ and $GT$ can be compressed. If this can be done then the schemes that have an element of $G2$ in the signature along with $G1$ can be further compressed.

%
%
%

\bibliographystyle{plain}
\bibliography{bibfile}
\section*{Appendix - Point compression and decompression code in Java}
\begin{lstlisting}[language=Java]
private static String pointcompress(Element elem) {
		// TODO Auto-generated method stub
		
		byte[] elembytes = elem.toBytes();
		int elemlength = elembytes.length;
		byte[] xbytes = new byte[elemlength / 2];
		byte[] ybytes = new byte[elemlength / 2];

		for (int i = 0; i < elemlength; i++) {
			if (i < elemlength / 2) {
				xbytes[i] = elembytes[i];
			} else {
				ybytes[i - (elemlength / 2)] = elembytes[i];
			}
		}
		
		BigInteger TWO = new BigInteger("2");
		BigInteger Xcord = new BigInteger(1,xbytes);
		BigInteger Ycord = new BigInteger(1,ybytes);
		
		String prefix = null;
		if((Ycord.mod(TWO)).compareTo(BigInteger.ZERO) == 0) {
			prefix = "2";
		}
		else {
			prefix = "3";
		}
		String XcordHex = Xcord.toString(16);
		String returnVal = prefix.concat(XcordHex).trim();		
		
		return returnVal;
	}
	
	
	
	private static Element pointdecompress(String compressedpoint, BigInteger prime, Field Group, BigInteger a,
			BigInteger b) {
		// TODO Auto-generated method stub

		String firstbyte = compressedpoint.substring(0, 1);
		String X = compressedpoint.substring(1, compressedpoint.length());
		System.out.println(X);
		BigInteger TWO = new BigInteger("2");

		BigInteger flag = new BigInteger(firstbyte, 16).mod(TWO);
		BigInteger Xcord = new BigInteger(X, 16);
		BigInteger YcordFinal = BigInteger.ZERO;

		BigInteger xpow3 = (Xcord.pow(3)).mod(prime);
		BigInteger ax = a.multiply(Xcord);

		int checklength = Group.newRandomElement().getImmutable().toBytes().length;

		BigInteger Ysquare = (xpow3.add(ax).add(b)).mod(prime);
		
		BigInteger Ycord = BigInteger.ZERO;

		// only works for curves where prime mod 4 = 3

		if(prime.mod(new BigInteger("4")).toString().equalsIgnoreCase("3")) {
			BigInteger pplus1by4 = (prime.add(BigInteger.ONE)).divide(new BigInteger("4"));
			Ycord = Ysquare.modPow(pplus1by4, prime);
		}
		// if prime mod 4 = 1
		else {
			Ycord = sqrtP(Ysquare, prime);
		}

		BigInteger modval = Ycord.mod(TWO);

		if (flag.compareTo(modval) == 0) {
			YcordFinal = Ycord;
		} else {
			YcordFinal = prime.subtract(Ycord);
		}


		byte[] xba = bigIntegerToBytes(Xcord);
		byte[] yba = bigIntegerToBytes(YcordFinal);

		byte[] xfinal = new byte[checklength / 2];

		byte[] yfinal = new byte[checklength / 2];

		if (xba.length < checklength / 2) {
			xfinal[0] = 0x00;

			for (int i = 0; i < xba.length; i++) {
				xfinal[i + 1] = xba[i];
			}
		} else {
			System.arraycopy(xba, 0, xfinal, 0, xba.length);
		}

		if (yba.length < checklength / 2) {
			yfinal[0] = 0x00;

			for (int i = 0; i < yba.length; i++) {
				yfinal[i + 1] = yba[i];
			}
		} else {
			System.arraycopy(yba, 0, yfinal, 0, yba.length);
		}

		byte[] c = new byte[xfinal.length + yfinal.length];

		System.arraycopy(xfinal, 0, c, 0, xfinal.length);
		System.arraycopy(yfinal, 0, c, xfinal.length, yfinal.length);

		Element sig = Group.newZeroElement();
		int sign = sig.setFromBytes(c);
		return sig.getImmutable();
	}
\end{lstlisting}
\end{document}